\documentclass[11pt]{article}
\usepackage{titlesec}

\titleformat{\section}
{\normalfont\Large\bfseries}
{\thesection.}
{1em}
{}
\usepackage{amssymb}
\usepackage{amstext}
\usepackage{latexsym}
\usepackage{graphicx}
\usepackage{amsmath}

\title{$\,$\\[-12ex]
	$\,$\hspace*{6.40in}\smash{\IfFileExists{hebrew.eps}{\includegraphics{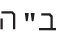}}{}}\hspace*{-200ex}\\[1ex]
	{\bfseries\huge 
		The Concept of a $J$-string 
}}

\author{ {\bf\Large Aharon Nudelman${}^*$}\\[.75ex]
	P.O.\ Box 7377\\
	Ashkelon, 78172 Israel\\
	E-mail: {\tt nudelman@ece.ucsd.edu}
}

\date{}

\newcommand{\eq}[1]{(\ref{#1})}

\newcommand{\dfn}{\sffamily\slshape}
\newcommand{\subt}{\bfseries\slshape}
\renewcommand{\em}{\dfn}

\newcommand{\LV}{\mbox{VL}}

\makeatletter
\newcommand\Isubsubsection{\@startsection{subsection}{2}{\z@}%
	{-1.25ex\@plus -1ex \@minus -.2ex}%
	{0ex \@plus .2ex}%
	{\normalfont\normalsize\bfseries}}
\makeatother

\makeatletter
\renewcommand\subsubsection{\@startsection{subsubsection}{3}{\z@}%
	{-1.25ex\@plus -1ex \@minus -.2ex}%
	{0ex \@plus .2ex}%
	{\normalfont\normalsize\bfseries}}
\makeatother

\newcommand{\sadvance}%
{\addtocounter{subsection}{1}\setcounter{subsubsection}{0}\vspace{1.75ex}}

\newcommand{\sreset}{\setcounter{subsection}{1}\setcounter{subsubsection}{0}}

\mathchardef\inn="3232

\begin{document}
	

	\maketitle
	\vspace{.250cm}
	
	\hfuzz=0pt
	
	
	\begin{center}
		{\bf Abstract} \vspace{-1.750ex}
	\end{center}
	\begin{list}{}
		{
			\addtolength{\leftmargin}{-1.9ex}
			\setlength{\rightmargin}{\leftmargin}
		}
		\item\noindent
		\baselineskip2.7ex
		\small
		\looseness=-1
		Certain linear objects, termed {physical lines},
		are considered, and initial assumptions concerning their
		properties are introduced.  A physical line in the 
		form of a circle is called a \emph{J-string}.
		It is assumed that a~\mbox{$J$-string} has an angular 
		momentum whose value is $\hbar$.
		It is then established that a~$J$-string 
		of radius $R$ possesses a mass $m_{\!_J}$,
		equal to~$h/2\pi c R$, 
		a~corresponding energy, as well as a charge $q_{\!_J}$, where 
		$q_{\!_J} = (hc/2\pi)^{1/2}$.
		It is shown that this physical curve consists of indivisible 
		line segments of length $\ell_\Delta = 2\pi({\hbar G/c^3})^{1/2}$, 
		where $c$ is the speed of light and $G$ is the gravitational constant.
		Quantum features of $J$-strings are studied.
		Based upon investigation of the properties and 
		characteristics of $J$-strings, a~method is 
		developed for the computation of the Planck 
		length and mass $(\ell_P, m_P)$.
		The values of $\ell_P$ and~$m_P$ 
		are computed according to the resulting 
		formulae (and given in the paper).
	\end{list}
	
	\vfill

	

	
	\footnotetext[1]
	{\\
		${}^*$This work was carried out in part while
		the author was a doctoral student and a senior researcher
		at the CSREP \hspace*{1.25ex}Reseach Institute, Moscow, Russia. 
	}

	
	\thispagestyle{empty}

	\section{Introduction}\label{sec0}
	\sreset
	
	\setcounter{subsection}{0}
	\Isubsubsection{}\label{0.1.1}
	\looseness=-1
	String theory is one of the most active areas of research 
	in modern physics at the present time.  Major contributions to this 
	field were made by Green, Polchinski, Schwartz,
	Witten and others, see~\cite{DEFJKMMW,GSW,Polchinski}
	and references therein.
	As is well known, considerable difficulties
	invariably arise in the investigation of strings and in the 
	characterization of their properties.  Efforts to overcome these
	difficulties through extension and elaboration of the existing
	mathematical apparatus do not always lead to a resolution
	of the underlying problems on a~fundamental level.
	In the present work, we limit our consideration 
	to certain inherently physical aspects of string theory;
	in particular, we introduce a certain new physical
	approach to the investigation of strings.\pagebreak[3.99]   
	In formulating this approach, we have made use,
	to some extent, of the well-known model of de Broglie 
	introduced in 1923 --- namely, a~string in the form of a circle that
	may be characterized in the four-dimensional space-time.

	\Isubsubsection{}\label{0.1.2}
	Among the range of questions and problems 
	that arise in string theory, one can single out those 
	dealing with the Planck length and the Planck mass.
	The formulae currently in use for the computation
	of these fundamental quantities (as well as the Planck time) were 
	derived by Planck back in~1899, by means of an analysis of the units of
	measure of three primary constants:~$h$,~$c$, and $G$.
	However, the ``natural'' (objective) physical units
	introduced by Planck have gained widespread notoriety
	only much later (it was not before the 1950's
	that there was active interest in the Planck length;
	see, for example~\cite{Wheeler}).
	As is well known, questions concerning the Planck
	length constitute one of the key problems in 
	string theory today (see the lecture notes~\cite[\mbox{pp.\,5--7}]{Greene}, 
	for example).
	Thus one cannot accept as satisfactory a situation where
	the values of these fundamental physical quantities
	are computed solely from the analysis of units of measure
	of certain constants. In this work, based upon the 
	aforementioned approach to the study of strings, 
	we propose an~alternative method for the computation
	of the Planck length and the Planck mass, and develop
	the corresponding formulae. 
	Since this 
	article is intended as a discussion item, 
	certain key issues and concepts are described in some detail. 
	The article is subdivided into individually numbered paragraphs
	(we use this numbering for internal reference).

	\vfill
	
	\vspace{4.5ex}
	\section{Initial assumptions}\label{sec1}\vspace{-1ex}
	\sreset
	
	\subsubsection{}\label{1.1.1}
	To quote~\cite{Einstein}, 
	a fundamental postulate of the special theory of 
	relativity is formulated as~follows:    
	\begin{list}{}
		{
			\addtolength{\leftmargin}{+5ex}
			\setlength{\rightmargin}{\leftmargin}
		}
		\item\noindent
		\sf
		Light is always propagated in empty space with a definite speed $c$
		that is independent of the state of motion of the emitting body.
	\end{list}
	The foregoing postulate, in essence, consists of two independent 
	assertions:
	\vspace{-1ex}
	\begin{list}{}
		{
			\addtolength{\leftmargin}{+5ex}
			\setlength{\rightmargin}{\leftmargin}
		}
		\item[{\bf 1.}]\noindent
		Light always has a certain nonzero velocity. 
		This means that the quanta of light (photons) are 
		always\,---\,that is, in any frame of reference\,---\,in motion.
		\item[{\bf 2.}]\noindent
		The value $v$ of the velocity of light (photons) is 
		always\,---\,that is, in any frame of reference\,---\,constant 
		and equal to $c$.
	\end{list}
	
	\subsubsection{}\label{1.1.2}
	\looseness=-1
	It is inherent to 
	this postulate that the first assertion above does not require indication of 
	a frame of reference relative to which the motion of photons 
	is considered. Consequently, there exist physical objects 
	for which the assertion that they are in motion is true without 
	specifying an observation framework, that is, without 
	choosing a definite frame of reference.
	
	\subsubsection{}\label{1.1.3}
	For any physical object moving at a certain velocity $v$, 
	the dimension $L_v$ along the direction of its motion,
	according to Lorentz transformations, is given by
	$$
	L_v \ = \ L_0 \displaystyle \sqrt{1 - \frac{v^2}{c^2}}
	$$
	where $L_0$ is the dimension of this object at rest.
	If $v \to c$, then $L_v \to 0$. Let us adopt~the~following 
	assumption: at $v = c$, the value of $L_v$ may be 
	actually equal to zero.
	
	\newpage
	\subsubsection{}\label{1.1.4}
	Let us introduce a postulate, termed the \emph{{\LV}-postulate},
	that relates ultimate velocity to vanishing (zero) dimension.\vspace{-1ex} 
	\begin{list}{}
		{
			\addtolength{\leftmargin}{+0ex}
			\setlength{\rightmargin}{\leftmargin}
		}
		\item\noindent
		\sf
		If the dimension of a physical object is equal to zero
		along a given straight line (direction), then\,---\,assuming conservation
		of energy of the object\,---\,this is necessary and sufficient 
		for this object to move in vacuum at an ultimate velocity
		along the given line (direction).\vspace{-2.5ex}
	\end{list}
	
	\sadvance
	\subsubsection{}\label{1.2.1}
	\looseness=-1
	Let us assume the following: vacuum is an aggregate of individual
	\emph{physical points}.~Using the general approach of 
	Cantor~\cite{Cantor} to the description of 
	``aggregates'' (manifolds) of mathematical points,
	physical points can be defined
	as objects of zero energy that have zero dimension
	in every direction.
	
	\subsubsection{}\label{1.2.2}
	For an individual physical point, it becomes necessary to observe that,
	according to the {\LV}-postulate, it is always in motion at the ultimate
	velocity simultaneously in {all\/} directions.
	This means that no displacement of the physical point takes place.

	\subsubsection{}\label{1.2.3}
	For future reference to the foregoing notion of vacuum,
	let us introduce the terminology \emph{physical vacuum} 
	to denote a vacuum as it is defined in \S\ref{1.2.1} and \S\ref{1.2.2}.

	\sadvance
	\subsubsection{}\label{1.3.1}
	Let us suppose that physical vacuum may contain, in addition to
	individual and independent physical points, also associations 
	of such points that are bound together in some manner.
	Let us assume that the foregoing associations of points are characterized 
	by local continuity and are always linear. We call such associations 
	of ponts \emph{physical lines}. Let us also assume that there
	exist physical lines that are characterized by both local continuity 
	and local curvature; 
	we call this type of physical lines \emph{physical curves}.

	\subsubsection{}\label{1.3.2}
	A continuous physical line of a certain length $L$
	cannot be partitioned, ``dissected,'' into separate
	physical points: any part of this line, 
	as small as desired, is still a line segment 
	and not a point.
	Let us assume that there has to exist a special physical
	object --- a certain small line segment 
	that is indivisible into parts.
	
	\subsubsection{}\label{1.3.3}
	Let us, for the time being, denote the length of this object by 
	$z_{\Delta}$. We call this object \emph{a~physical line-element}, 
	and define it as follows.\vspace{-1.25ex}
	\begin{list}{}
		{
			\addtolength{\leftmargin}{+3ex}
			\setlength{\rightmargin}{\leftmargin}
		}
		\item\noindent
		\sf
		A physical line-element is a continuous linear association 
		of physical points, indivisible into parts, that has length  
		$z_{\Delta}$ and zero cross-section.\vspace{-.75ex} 
	\end{list}
	Let us assume that the properties of this physical
	object can be characterized in two ways.
	The object can be interpreted as a single indivisible 
	conglomerate of physical points and, at the same time,
	as a~free assemblage thereof.
	
	\subsubsection{}\label{1.3.4}
	A physical line of length $L$, as well as any other line, 
	can be either continuous throughout or discrete --- that is,
	constructed of continuous segments and gaps between them.
	The length~$l$ of any segment (or gap) must be an
	integer multiple of $z_{\Delta}$.
	
	\sadvance
	\subsubsection{}\label{1.4.4}
	Let us introduce the following working hypothesis:\vspace{-1.25ex}
	\begin{center}
		\sf
		In the observable Universe, there exists nothing
		except physical vacuum and physical lines.
		\vspace{-.75ex}
	\end{center}
	
	\subsubsection{}\label{1.4.5}
	Let us point out the main consequence of the foregoing 
	hypothesis.\vspace{-1.25ex}
	\begin{center}
		\sf
		All the energy and mass in the manifest Universe
		are concentrated in~the form of physical lines.
		\vspace*{-3ex}
	\end{center}

	\newpage
	\vspace{4.5ex}
	\section{%
		Planar {\kern -4pt} 
		physical {\kern -4pt} 
		lines {\kern -4pt} 
		of {\kern -4pt} 
		constant {\kern -4pt} 
		curvature {\kern -4pt} 
		and {\kern -4pt} 
		their {\kern -4pt} 
		\mbox{properties}}
\label{sec2}\vspace{-1ex}
\sreset

\subsubsection{}\label{2.1.1}
\looseness=-1
Consider the following hypothetical object: 
a special continuous closed physical curve in the form of a circle.
Let us call 
this physical curve a~\mbox{\emph{$J$-string}}.
We adopt the following postulate that may be regarded
as the definition~of~a~\mbox{$J$-string}.
\vspace{-1ex} 
\begin{list}{}
	{
		\addtolength{\leftmargin}{+3ex}
		\setlength{\rightmargin}{\leftmargin}
	}
	\item\noindent
	\sf
	A $J$-string is a special continuous closed physical curve 
	in the form of a circle that constitutes an association
	of physical line-elements and, at the same time,
	the trajectory along which these line-lements are moving
	so that the angular momentum of a~$J$-string is
	equal to $\hbar$.
\end{list}

\subsubsection{}\label{2.1.2}
Given a physical curve, we denote the length, as measured along the arc,
of a constituent physical line-element by $\ell_{\Delta}$.
Let us now investigate the physical properties of
a line-element of length $\ell_{\Delta}$ and of
a~$J$-string.
Consider an arbitrary stationary (non-rotating) circle. Let us distinguish a small segment of 
length~$\Delta L$ on the circle, and draw a tangent through one of 
its points. The dimension of the segment along
this tangent is equal to the length its projection 
onto the tangent.

By definition, a $J$-string is an association of physical line-elements 
jointly comprising a circle. Let us now draw tangents to one 
such line-element --- a circular segment of length $\ell_{\Delta}$. 
Assume, hypothetically, that tangents are drawn through 
each point of this line-element. It would appear that the dimension
of the line-element along any of these tangents must be also equal 
to the length of its projection onto the corresponding tangent.

\looseness=-1
According to \S\ref{2.1.1}, all the line-elements of a~$J$-string are 
moving in a circular trajectory, which coincides with the $J$-string itself.
The linear velocity $v$, at each point of any such line-element 
of length $\ell_{\Delta}$, is directed along a tangent to the trajectory.
One can assert the following:
any such line-element is moving with linear velocity $v$
simultaneously along all tangents to the element.

At a given moment $t$, each of the points of the line-element is moving,
with velocity $v$, solely along the tangent 
corresponding to this point: point $A$ is moving along the tangent 
at~$A$, point~$B$ is moving along the tangent at~$B$, and so forth. 
It follows that at time $t$, the dimension of the line-element 
along any given tangent (for example, along the tangent at $A$) 
is equal to the ``length'' of the point of tangency (point $A$ itself) ---
that is, equal to zero. Invoking the {\LV}-postulate,
we come to the conclusion that the linear velocity $v$ along any tangent 
to the considered line-element must necessarily be ultimate: $v = c$. 
The~foregoing analysis applies to each
of the line-elements comprising a $J$-string; hence, it also
holds for the $J$-string as a whole.\vspace{1ex}

\looseness=-1
{\subt A comparative physical model.}  
Consider a closed non-elastic (``soft'') thread of a certain mass 
that has an arbitrary shape and lies on a planar horizontal surface.
Suppose that the surface and the thread are rotated 
in such a manner that the thread takes the shape of 
a ring of radius~$R$, uniformly distended along all of its
length by a~radial centrifugal force (assume that the ring 
is rotating in its plane about its geometric center).
Let us think of the rotating ring as a torus,
and let us call the circular line centrally located within the 
torus the {\sl circular axis of the ring}. 
The linear velocity of rotation is then directed along
a tangent to the circular axis of the ring (torus) and has
some arbitrary value~$v$. 
The diameter of the cross-section of the ring (torus) is equal to $d$, 
where $d \ll R$. 

\looseness=-1
Each small segment of the ring will have a certain nonzero dimension 
along the direction of its velocity --- that is, 
a certain non-vanishing projection onto any tangent to the  
circular axis of the ring (torus). The length of the projection,
onto tangent $A$ for example, is equal to the length (or part 
of the length) of the long diameter of the ellipse 
that arises as the cross-section of the ring (torus) by a plane
perpendicular to the horizontal surface and passing\pagebreak[3.99] 
through the point of tangency $A$.
Now suppose that $d \to 0$.
The physical situation described above remains the same.
However, if $d$ were actually equal to $0$, the situation
would have been fundamentally different.
The ellipse obtained by the aforementioned cross-section 
of the ring collapses into a point (the point of tangency).
In this case\,---\,according to the {\LV}-postulate\,---\,the 
linear velocity~$v$ is no longer arbitrary, it must necessarily 
have the ultimate value.

\sadvance

\subsubsection{}\label{2.1.3}
Let us draw, through each point of the physical line-element
considered in \S\ref{2.1.2}, a~plane perpendicular to the tangent
at that point. The size of the cross-section of a physical line 
by such a plane is clearly zero.
It follows that each line-element of a $J$-string, 
according to the~{\LV}-postulate, must 
move\,---\,along certain directions\,---\,in the plane 
of its cross-section, at the ultimate velocity.
In order to find these directions, let us again consider the
physical line-element of length $\ell_{\Delta}$ and draw 
tangents through all of its points.
In addition to the tangents,
let us draw through each point of the line-element 
two mutually perpendicular straight lines (normals) that lie
in the plane of the cross-section. 
Thus at any point of the line-element,
a~tangent and the two normals form three orthogonal axes.

\subsubsection{}\label{2.1.4}
\looseness=-1
Let us assume that one of the two normals lies in the plane of 
the $J$-string that contains the considered line-element.
This normal will be directed radially, either from the center
or towards the center of the $J$-string. According to the
{\LV}-postulate, all the points of the considered line-element 
(and, hence, the entire element) must move at the ultimate velocity 
simultaneously along all radial directions. Such simultaneous 
motion of the line-elements in radial directions
corresponds to a concentric expansion (or concentric contraction)
of the entire $J$-string, while the geometric center of the 
$J$-string remains at rest. Thus we have a resting~$J$-string
that expands (or contracts) radially with velocity $c$. 
Let us call such a state of the $J$-string \emph{state $\alpha$}.

\subsubsection{}\label{2.1.5}
If the first of the two normals under consideration lies in the plane 
of the $J$-string, then the second normal is perpendicular to this plane. 
Each point of the physical line-element under consideration and,
hence, the entire element (as well as all the other elements of 
the \mbox{$J$-string}), must necessarily move at the ultimate velocity 
along the second normal, in accordance with the~{\LV}-postulate.
This means that the geometric center of the $J$-string must move 
at the ultimate velocity $(v = c)$ along an axis perpendicular to 
the plane of the $J$-string. (Along such an axis, not only every
line-element of the circle, but also the $J$-string as a whole, 
has dimension zero.) Thus we have a~$J$-string that
moves along a line perpendicular to its plane with velocity~$c$. 
Let us call such a state of the $J$-string \emph{state $\beta$}.

\subsubsection{}\label{2.1.6}
\looseness=-1
Consider again the line-element of~\S\ref{2.1.2}.
Let us fix an arbitrary plane of its cross-section and draw 
an arbitrary straight line in this plane.
Assume that this line passes through the section point and is inclined 
at an angle~$\theta$, where $0 < \theta < \pi/2$, to the plane of 
the $J$-string. Furthermore, let us draw lines parallel to the 
line described above through all the points of the line-element.

In our analysis of states $\alpha$ and $\beta$, we have considered 
the motion of each point of the physical line-element\,---\,at velocity~$c$,
along a given axis\,---\,as being independent from the motion of all
other points, as long as the integrity of the line-element was 
maintained (such analysis is based upon the properties
of a physical line-element, described in~\S\ref{1.3.3}).
However, the case considered here is different from the two 
cases described in the foregoing paragraphs. Any specific point 
of the line-element and, hence, the element as a 
\mbox{whole\,---\,as a~separate entity\,---\,can}
move at the ultimate velocity, at least
in principle, along an arbitrarily straight line (and lines parallel
to this straight line),
such as the one(s) described above. But for a~$J$-string as~a~whole
such motion is impossible. Unlike a physical line-element,
which by definition ({cf}.~\S\ref{1.3.3}) is an association of points,
\mbox{a~$J$-string} is an association of line segments, that is,
physical line-elements of length~$\ell_{\Delta}$.
Hence a~$J$-string
always has a certain nonzero dimension along a line inclined at
an angle~$\theta$ to its plane --- the projection of a $J$-string, as a whole,
onto any such line is not equal to zero.

\looseness=-1
It follows that in this physical situation, the {\LV}-postulate
does not apply. Consequently, the two assertions of the {\LV}-postulate,
about the necessity of motion and about the ultimate velocity of motion,
do not extend to the case considered here. In this work, we assume that 
motion of a~$J$-string with velocity $v$, where $0 \le v < c$,
along an axis inclined at an arbitrary angle $\theta \ne \pi/2$
to its plane is possible but not necessary. We say 
that a~$J$-string in such a motion is in \emph{state $\gamma$}.

\subsubsection{}
Assume that if certain conditions arise, a $J$-string can 
transit from one state to~another.

\subsubsection{}\label{2.1.7}
\looseness=-1
In addition to the analysis carried out 
in~\S\ref{2.1.2}\,--\,\S\ref{2.1.5}, it 
is essential to point out the following with
respect to the motion at ultimate velocity
considered in these paragraphs. For example, as shown in~\S\ref{2.1.5},
a~$J$-string in state~$\beta$ must move at velocity $v = c$
along an~axis perpendicular to its plane. The motion along this axis 
can occur in any of two opposite directions. If there 
is no ``preferable direction'' (that is, if the two directions
are completely equivalent --- there is nothing that 
may distinguish one direction relative to the other),
then an actual displacement of the $J$-string along this axis
is impossible. According to the~{\LV}-postulate, the~$J$-string will 
move at ultimate velocity {simultaneously\/} in two opposite 
directions (see, for comparison,~\S\ref{1.2.2}).
This condition\,---\,that there exist a  ``preferable
direction'' for an~actual displacement to occur\,---\,clearly extends
to the circular rotation of a $J$-string, considered in~\S\ref{2.1.2},
and also to the potential expansion or contraction of a~$J$-string
in state~$\alpha$, considered in \S\ref{2.1.4}. In~this~regard,
we point out the following:\vspace{-6.0ex}
\begin{list}{}
	{
		\addtolength{\leftmargin}{-1ex}
		\setlength{\rightmargin}{\leftmargin}
	}
	\item\noindent
	\begin{description}
		\item[For state $\alpha$:]
		Conditions for the realization by a~$J$-string
		of its potential concentric expansion or contraction 
		are outlined in~Section\,\ref{sec3} of this 
		paper (see~\S\ref{4.2.1}). 
		
		\item[For state $\beta$:] 
		Investigation of the physical factors that may cause a~$J$-string 
		to move in one of the two directions along an axis perpendicular 
		to its plane is beyond the scope of this work.\vspace{1ex} 
		
		\item[For circular rotation:] $\,\hfill$
		
		\begin{itemize}
			\item 
			The movement (displacement) of the line-elements comprising a $J$-string 
			along a circular trajectory is a fundamental physical property of 
			the~$J$-string;  
			this property follows directly from the condition that a $J$-string 
			has to have an angular momentum (see the definition in~\S\ref{2.1.1}).
			
			\item 
			Any physical object has an angular momentum if and only if
			a preferable direction of rotation is given in some manner.
			\vspace{-1.0ex}
		\end{itemize}
	\end{description}
\end{list}

\vspace{3.5ex}
\section{Parameters and characteristics of\\
	planar physical lines of constant curvature\hspace*{5ex}}
\label{secnew}\vspace{-1.25ex}
\sreset

\subsubsection{}\label{2.2.1}
\looseness=-1
Let us consider a $J$-string of radius $R$ in state $\gamma$. 
Suppose that this $J$-string is moving (actual displacement takes
place) in a certain frame of reference at velocity $v$, 
where $v \to 0$. Let us now assume that 
the value of $v$ becomes zero.
By definition, any~$J$-string is a continuous physical line 
in the form of a circle. If~$v = 0$, then a single parameter
is necessary and sufficient to completely characterize 
the~$J$-ob\-ject: its radius~$R$. 
Based on~\S\ref{1.4.5}, let us assume that the $J$-string must have a 
{\sl mass}, 
which we denote by $m_{\!_J}$. It~follows that the value of $m_{\!_J}$ 
depends {\sl only} on the radius of the~$J$-string,
that is $m_{\!_J} = f(R)$.\pagebreak[3.99]

\subsubsection{}\label{2.2.2}
In~\S\ref{2.1.2}, we have considered, as a comparative physical model,
a~thread whose cross-section has diameter $d \ne 0$. This object,
as is well known, consists of corpuscles (atoms and molecules)
bound together in some manner. Let us call this and similar objects 
{\em corpuscular lines\/}.

Suppose that a corpuscular line of mass $m$, in the 
shape of a circle of radius $R$, rotates in 
its plane about its geometric center. Let $T$ denote 
the magnitude of the angular momentum of this corpuscular line.
It is well known that\vspace{-1ex}
$$
T \ = \ m v R
\vspace{-.5ex}
$$
where $v$ is the magnitude of the linear velocity of rotation,
directed along a tangent to the circle.
Let us assume that this expression also 
holds for a~$J$-string of radius $R$.
As~shown~in~\S\ref{2.1.2} above, for a~$J$-string we have $v = c$,
and consequently\vspace{-.5ex}
\begin{equation}
	\label{e1}
	T_{\!_J} \ = \ m_{\!_J} c R
	\vspace{-1.5ex}
\end{equation}
According to 
\S\ref{2.1.1}, we have\vspace{-1.0ex}
\begin{equation}
	\label{e2}
	T_{\!_J} \ = \ \frac{h}{2\pi}
\end{equation}
From~\eq{e1} and~\eq{e2}, we have
\begin{equation}
	\label{new3}
	m_{\!_J} c R \ = \ \frac{h}{2\pi}
\end{equation}
Thus we arrive at the following expression for the function $f(R)$ 
that was introduced in the foregoing paragraph:\vspace{-1.0ex}
\begin{equation}
	\label{e4}
	m_{\!_J} 
	\, = \, \frac{h}{2 \pi c} \, \frac{1}{R}
	\vspace{1ex}
\end{equation}

\subsubsection{}\label{2.2.3}
The foregoing relation between the mass of a $J$-string 
and its radius may be also deduced in an alternative manner,
using special theory of relativity. 
Assume that the $J$-string in\pagebreak[3.99]
state~$\gamma$ is moving (actual
displacement takes place) in its own plane with velocity $v$,
while maintaining the shape of a circle; the velocity vector lies 
in the plane of the~$J$-string, and $0 \le v < c$ 
(the preservation of the circular shape by the $J$-string is proved in \S6.3.1 and \S6.3.2). 
One can write-down the following well-known relations
for the parameters of this $J$-string:
$$
R_v \ = \ 
R_0 \displaystyle{\sqrt{1 - \frac{v^2}{c^2}}}
\qquad\ \  \text{and} \qquad
m_{\!_{J,{\scriptstyle v}}} 
\ = \
m_{\!_{J,0}} \displaystyle{1 \over \sqrt{1 - \frac{v^2}{c^2}}}
\vspace{-1.0ex}
$$
from which it follows that 
$$
\displaystyle\frac{R_v}{R_0} 
\ = \ 
\displaystyle{\sqrt{1 - \frac{v^2}{c^2}}} 
\ = \
\displaystyle\frac{m_{\!_{J,0}}}{m_{\!_{J,{\scriptstyle v}}}} 
$$
Since $m_{\!_J} = f(R)$ as shown in~\S\ref{2.2.1},
we have $R_v/R_0 = f(R_0)/f(R_v)$.
This means that for any value of the velocity $v$,
the following relation holds:\vspace{-.50ex}
$$
R_0 f(R_0) \,=\, R_v f(R_v) \,=\, \text{const} 
\vspace{-.50ex}
$$
Denoting the value of the constant above by $b$, 
we obtain the expression 
\begin{equation}
	\label{e5}
	f(R)  \ = \  b\,\frac{1}{R} \ = \ m_{\!_J}
\end{equation}
where $b$ must be equal to $h/2\pi c$ according to~\eq{e4}.

\sadvance
\subsubsection{}\label{2.3.1}
As is well known, if a material point is moving in a circular 
trajectory, then the frequency of its oscillations with respect to 
any axis passing through a diameter of the trajectory\pagebreak[3.99]
is identically equal to the frequency of its revolutions along the 
circumference. 
Let us introduce the concept of \emph{frequency of a~$J$-string}
and denote it by $\nu_{\!_J}$. Let us adopt the following expression
for the frequency of a~$J$-string:\vspace{-.50ex}
\begin{equation}
	\label{e7}
	\nu_{\!_J}
	\ = \
	\frac{c}{2\pi R}
	\vspace{.50ex}
\end{equation}
The parameter $\nu_{\!_J}$ reflects the rotation of a~$J$-string in its 
plane, that is, the motion of each of its line-elements with velocity $c$ 
along the circle as a trajectory 
(as postulated in~\S\ref{2.1.1}).\pagebreak[3.99]

According to \S\ref{1.4.5}, a~$J$-string must have not only 
mass $m_{\!_J}$, but also {\sl energy}.
Starting with~\eq{new3}, we obtain the 
following relation
\begin{equation}
	\label{e8}
	m_{\!_J} c^2 R
	\ = \
	\frac{hc}{2\pi}
	\vspace{-1.5ex}
\end{equation}
It now follows from~\eq{e7} and \eq{e8} that
\begin{equation}
	\label{e9}
	m_{\!_J} c^2  
	\ = \
	h \nu_{\!_J}
	\vspace{.75ex}
\end{equation}
Suppose that a~$J$-string has energy associated with its rotation 
in its own plane. Let us denote this energy by $K_{\!_J}$. 
We adopt the following assumption: the expression $h\nu_{\!_J}$ can be
interpreted as a formula describing the energy $K_{\!_J}$, that is
\begin{equation}
	\label{e10}
	K_{\!_J} 
	\ = \
	h \nu_{\!_J}
	\vspace{-1.5ex}
\end{equation}
Then, in view of~\eq{e9}, we have
\begin{equation}
	\label{e11}
	K_{\!_J} 
	\ = \
	m_{\!_J} c^2  
	\vspace{.75ex}
\end{equation}
Let us {\sl formally\/} regard the energy $K_{\!_J}$ as the 
\emph{kinetic energy of a~$J$-string}.

\subsubsection{}\label{2.3.2}
\looseness=-1
The $J$-string under investigation in the foregoing paragraphs, 
according to \S\ref{2.2.1}, is in state $\gamma$ and its velocity 
is assumed to be zero. 
Assume, hypothetically, that for a short period of time 
$\Delta t$ the motion of the physical line-elements along the
circumference of the $J$-string has ceased.
In this (completely hypothetical) case, 
during the time~$\Delta t$, the $J$-string
would be a~static line in the form of a circle.
Such an object, in principle, cannot have kinetic energy.
However, a~static physical line in the form of
a circle still embodies a~certain type 
of geometric information, namely a certain well-defined 
order of a part of physical vacuum.
Let us assume that a~$J$-string has, 
in addition to the energy $K_{\!_J}$ 
associated with its rotation, also another kind 
of energy --- the energy $U_{\!_J}$ associated solely
with the shape and radius of its physical line.
Let us {\sl formally\/} regard the energy $U_{\!_J}$ as the 
\emph{potential energy of a~$J$-string}.

\vspace{4.5ex}
\section{$J$-strings}\label{sec3} \vspace{-1ex}
\sreset

\subsubsection{}\label{3.1.1}
Let us continue the investigation of a~$J$-string of radius $R$ 
and mass $m_{\!_J}$ in state~$\gamma$, under the condition that $v = 0$.
This physical object is a corporeal line of certain mass that has
the form of a circle and rotates in its plane. 
For all known objects of this type (for example,
for the corpuscular line considered in~\S\ref{2.2.2}),
each small line segment $\Delta L$ is subject to a centrifugal force
$
({\Delta L}/{2 \pi R}) \left( m v^2/{R} \right)
$.
Let us assume that the corresponding centrifugal force is
present in a~$J$-string as well, and that its value can 
be analogously described as
$
({\ell_{\Delta}}/{2 \pi R}) \left( m_{\!_J} c^2/{R} \right)
$.
Let us introduce a force $F_c$, 
regarded as a certain \emph{aggregate centrifugal force},
whose value is given by
\begin{equation}\label{e12}
	F_c
	\ = \ 
	m_{\!_J} \frac{c^2}{R}
\end{equation}

\subsubsection{}\label{3.1.2}
If $F_c$ were the only force acting on a~$J$-string, it
would tear apart. However, in this work we assume that
a~$J$-string is stable. Therefore, it is necessary to 
further assume the existence of a \emph{centripetal force} that 
is equal to $F_c$ in its magnitude and opposite to $F_c$
in its direction: this force is directed along 
the radius, towards the geometric center of a~$J$-string. 
Let us denote the centripetal force by $F_{\!_J}$. Then,
in view of~\eq{e12}, we have
\begin{equation}
	\label{e13}
	F_{\!_J}
	\, = \,
	F_c
	\ = \ 
	\frac{m_{\!_J}c^2}{R}
\end{equation}
(Here, $F_{\!_J}$ and $F_c$ denote the magnitudes of the two 
corresponding aggregate forces.)
One can assume that the energy $U_{\!_J}$, introduced in~\S\ref{2.3.2},
is associated with the action of the force $F_{\!_J}$.

\sadvance
\subsubsection{}\label{3.2.1}
In \S\ref{2.1.2} we have introduced and studied,
as a comparative physical model,
a corpuscular line in the shape of a circle that
is rotating in its plane with velocity $v$. 
The corpuscular line can be 
formally considered as a ``string'' (elastic cord)
that is uniformly distended by a centrifugal\pagebreak[3.99]
force along its perimeter.
Such a ``string'' is at rest --- there is no oscillation
(or rotation) of its line of circumference.
This interpretation of a rotating corporeal 
line is fully applicable to the object under investigation,
a physical curve of mass $m_{\!_J}$ rotating in its
own plane, as well.

However, there is an essential difference between the two linear
objects. The corporeal object described above is a corpuscular line
in the shape of a circle rotating in its plane; at the same time,
this object can be {\sl formally\/} considered as a ``string'' at rest
(distented by a centrifugal force). 
In the present work, we assume that the object under investigation
has the following special property. It is, equivalently
\begin{list}{}
	{
		\addtolength{\leftmargin}{-1ex}
		\setlength{\rightmargin}{\leftmargin}
	}
	\item
	\begin{itemize}
		\item 
		a line in the form of a circle of radius $R$,
		rotating in its plane with velocity $c$;\\
		\hspace*{-9.5ex}and
		\item 
		a ``resting string'' of the same form,
		distended by a force whose value is $m_{\!_J} c^2/R$
		(the force is directed along the radii, away from the center).
	\end{itemize}
\end{list}
\looseness=-1
The foregoing property of a~$J$-string means that 
it is endowed with a~{\sl physical dualism}.

It~manifests itself as two fundamentally different objects:
a~{\sl kinematic object\/} (a ``rotating line'') or a~{\sl static object\/} 
(a~``resting string'').
Consequently, all the features and characteristics of a given $J$-string
can be determined in their entirety based on either of the 
two models: the kinematic model (used earlier in Sections~\ref{sec2}
and~\ref{secnew}) or the static~one.

\subsubsection{}\label{3.2.2}
As mentioned in \S\ref{2.1.1} and \S\ref{2.2.2}, 
the object under investigation has an angular momentum $T_{\!_J}$. 
This characteristic belongs solely to the kinematic model
of a ``rotating line.''
A~static model of a ``resting string'' 
cannot have angular momentum.
However, from the concept of the physical dualism of a~$J$-string, 
introduced above, it follows that 
the static model must have 
a~certain physical characteristic
that is strictly equivalent to~$T_{\!_J}$.

\subsubsection{}\label{3.2.3}
Let us introduce the notion of \emph{charge of a~$J$-string},
defined as follows:
\vspace{-1ex} 
\begin{list}{}
	{
		\addtolength{\leftmargin}{-1ex}
		\setlength{\rightmargin}{\leftmargin}
	}
	\item\noindent
	\sf
	The charge of a $J$-string is a dynamical characteristic 
	of this object, regarded as a ``resting string,'' that 
	is equivalent to its kinematic characteristic --- the angular momentum.
\end{list}
Let us denote the charge of a $J$-string as $q_{\!_J}$.
As is well known, the speed of light in vacuum~$c$
may be also regarded as the electrodynamic constant;
let us assume that the {\sl square of $q_{\!_J}$ is equal 
	to $T_{\!_J}$ multiplied by the electrodynamic constant}.

\subsubsection{}
Thus we have
\begin{equation}
	\label{e16}
	q_{\!_J}^2 
	\ = \ 
	c T_{\!_J}
\end{equation}
Next, in view of~\eq{e2} and~\eq{e16}, we arrive 
at the following expression\vspace{-1ex}
\begin{equation}
	\label{e17}
	q_{\!_J} 
	\ = \
	\left({\frac{hc}{2\pi}}\right)^{1 \over 2}
	\vspace{+.75ex}
\end{equation}
For a~$J$-string of radius $R$, according to~\eq{e8}, 
we have
\begin{equation}
	\label{e18}
	q_{\!_J}^2 
	\ = \ 
	m_{\!_J} c^2 R
\end{equation}
Comparing the above expression for $q_{\!_J}^2$
to~\eq{e13}, we conclude that the force $F_{\!_J}$ 
is given by 
\begin{equation}
	\label{e19}
	F_{\!_J}
	\ = \ 
	\frac{q_{\!_J}^2}{R^2}
\end{equation}

\sadvance

\subsubsection{}\label{3.2.5}
Taking into account \eq{e17}, the expression~\eq{e19}
for the force $F_{\!_J}$ can be rewritten as follows:
\begin{equation}
	\label{e20}
	F_{\!_J} 
	\ = \
	\frac{hc}{2\pi} \, \frac{1}{R^2}
\end{equation}
\looseness=-1
(The same relation holds for the force $F_c$ as well.)
It is well known that one of the expressions for the 
constant of fine structure is $2\pi e^2/ hc$,
where $e$ denotes the elementary electrical charge.
The inverse expression ${hc}/{2\pi e^2}$, 
according to existing computations based on 
available experimental data, evaluates to about~$137$
(more precisely, $137.036\ldots$) in vacuum. Thus by~\eq{e20} we have
$$
F_{\!_J} 
\ \cong \
137\,\frac{e^2}{R^2}
\ = \
137 F_e
$$
where $F_e$ is the magnitude of the force of interaction between 
two elementary electrical charges whose geometric centers are 
located at distance $R$ from each other.
Consequently, we have
$$
q_{\!_J}^2
\ \cong \
137 e^2
$$

\subsubsection{}\label{4.2.1}
Suppose we have a $J$-string in state $\alpha$, 
as described in \S\ref{2.1.4}.
In state~$\alpha$\,---\,according 
to the {\LV}-postulate\,---\, a~$J$-string must either concentrically 
expand or concentrically contract at ultimate velocity. 
Concentric contraction of an~isolated $J$-string 
is impossible. An isolated \mbox{$J$-string} of radius~$R$ and 
mass~$m_{\!_J}$\,---\,that is equal, in view of~\eq{e5},
to $b/{R}$\,---\,cannot 
give rise to \mbox{a~$J$-string} of a smaller 
radius $R - \Delta R$ and correspondingly higher mass
$b/({R \,{-}\,\Delta R})$. Consequently, an isolated 
$J$-string can (and does) only expand with velocity~$c$
(in-depth investigation of the concentric expansion of 
a~$J$-string is beyond the scope of the present work).

\vfill
\vspace{3.5ex}
\section{Quantization in $J$-strings}
\label{sec4}\vspace{-1ex}
\sreset

\subsubsection{}\label{4.1.1}
According to \S\ref{1.3.3} and \S\ref{2.1.2}, 
a physical line-element is an object of length $\ell_{\Delta}$ 
that is indivisible into parts. In this work,
it is assumed that $\ell_{\Delta}$ constitutes a~certain
fundamental length. Let us determine its value.
In particular, let us consider the possibility 
of identifying $\ell_{\Delta}$ with a~well-known
physical quantity --- the Plank length $\ell_{P}$.

\subsubsection{}\label{4.1.2real}
\looseness=-1
Let us investigate the statement: {\sl In quantum (discrete) space,
	any linear dimension is an integer multiple of the 
	fundamental length~$\ell_{P}$}. Let us proceed with
the following analysis. Consider, for example, 
a circle of radius~$R$. If $R$ is
an integer multiple of a~certain minimal possible length~$A$,
then $R = nA$ while the length $L$ of the circle itself 
is given by~$2\pi(nA) = n(2\pi A)$. This means that $L$ is an 
integer multiple of another minimal length, say $B$.
The ratio of the two indivisible ``units of length'' 
is $B/A = 2\pi/k$, where $k$ is an arbitrary positive integer 
(e.g., $k=1$)
or a~simple fraction. This ratio is 
irrational, and hence the lengths $B$ and $A$, in principle,
cannot be co-measurable. In other words, these lengths constitute 
completely independent linear parameters.
This leads to the conclusion that there should exist, not one,
but {\sl two fundamental~lengths}. 

\subsubsection{}\label{6.1.3}
Among the range of problems in string theory, the 
questions about the Planck length stand out as one
of the key issues in the field. It is well known
that the value of the Planck length is computed according 
to the formula $(\hbar G/c^3)^{1/2}$ and is equal
to $1.616{\cdot}10^{-35}{\rm m}$. At the present time,
the assumption that $\ell_P$ is a fundamental length,
which determines the dimensions of a quantum of space, 
is widely accepted. The foregoing analysis in~\S\ref{4.1.2real}
indicates that it would be necessary to consider, not one,
but two physical quantities associated with length: say $(\ell_P^*)_1$
and $(\ell_P^*)_2$; let us assume that $(\ell_P^*)_1 = A$ and
$(\ell_P^*)_2 = B$.

\subsubsection{}\label{6.1.4}
\looseness=-1
Let us extend the problem posed in~\S\ref{4.1.1}. 
The length of an element of the physical line 
in the form of a circle (that is, $\ell_{\Delta}$) is,
according to~\S\ref{4.1.2real}, the fundamental length $B$.
Thus we need to determine not only the value of $\ell_{\Delta}$,
but also the value of the fundamental length associated with the
radius of the circle --- that is, the length~$A$.

\sadvance
\subsubsection{}\label{4.1.2}
To determine the general form of the expressions for $A$ 
and $B$, we will employ certain assumptions concerning the
gravitational constant $G$, discussed later.
First, let us write the 
well-known expression for the interaction of 
the electrical charges of an electron and a~positron 
at rest (we assume that their centers of mass are 
at distance $L$ from each other) in a form that is formally
analogous to Newton's law, namely\vspace{.50ex}
$$
F_e = \frac{e {\cdot} e}{L^2} 
\ = \
\frac{
	\left(\sqrt{\varphi_0}m_0\right)
	{\cdot}
	\left(\sqrt{\varphi_0}m_0\right)}
{L^2}
\ = \ 
\varphi_0 \frac{m_0{\cdot}m_0}{L^2} 
\vspace{.750ex}
$$
where $\varphi_0 = (e/m_0)^2$ and $m_0$ is the mass of each of
the two particles,
assumed to be at rest in some frame of reference.
Let us compare this to Newton's law for the interaction of 
the masses of these same particles, namely $G(m_0{\cdot}m_0)/L^2$.
As is well known
$$ 
G 
\ \ll \
\left(\frac{e}{m_0}\right)^2
\ \qquad
\text{that is,}\ \
\varphi_0 \, \gg \, G
\vspace{.50ex}
$$ 

Let us now set both particles in parallel motion at velocity~$v$, in such 
a~way that the distance $L$ between them remains constant.\pagebreak[3.99]
Let $\varphi$ denote the square of the ratio of the charge
of each particle to its mass.
When $v$ approaches $c$, the mass $m_v$ of each of the two particles 
continuously increases while the value $\varphi$, correspondingly,
decreases (since $e = \text{const}$).
Suppose that $\varphi$ cannot decrease without bound.
Consequently, there has to exist a certain {\sl limiting value\/}
(let us denote it by~$\varphi_z$), to which $\varphi$, that is,
the square of the ratio of the charge of a particle to its mass,
must converge for $v \to c$.
Let us write this as follows
\begin{equation}\label{phi-limit}
	\lim_{v \to c}\varphi 
	\ = \ 
	\varphi_z
	\vspace{1.00ex}
\end{equation}
The above expressions 
are fully applicable to an individual, isolated,
electron (or positron) as well.
If this isolated particle is at rest in a certain frame of reference,
then $\varphi_0 = ({e}/{m_0})^2$, while if the particle is moving at 
$v \to c$, then, in view of~\eq{phi-limit}, 
\begin{equation}
	\label{last!}
	\lim_{v \to c} \left({e \over m_v}\right)^2  = \varphi_z
\end{equation}

\subsubsection{}\label{6.2.2}
Let us now consider $J$-strings. Assume that a~$J$-string 
of radius~$R$ is in state~$\gamma$ and that $v = 0$.
Such a $J$-string has charge $q_{\!_J}$ and mass $m_{\!_J}$, 
and we can write $\varphi_0^* = (q_{\!_J}/m_{\!_J})^2$.
Let us set the $J$-string in motion at velocity $v \to c$, 
in such a way that the velocity vector lies in the plane 
of the $J$-string. As shown in~\S\ref{2.2.3}, the mass 
of such an object (which retains the shape of a circle) continuously increases with its velocity.
Since the charge $q_{\!_J}$ of a~$J$-string is constant, it follows 
that when $v \to c$ the square of the charge-to-mass 
ratio\,--\,that is, the value $\varphi^*$\,--\,continuously 
decreases. One may assume that in this case as 
well (that is, for a~$J$-string) there exists a certain 
limiting value $\varphi_z^*$; 
we write this as 
$$
\lim_{v \to c}\varphi^* = \varphi_z^* 
$$

\subsubsection{}\label{6.2.3}
\looseness=-1
Consider a sequence of $J$-strings with decreasing radii 
$R_1 > R_2 > R_3 > \cdots\ $. 
The length of the circumference 
of a~$J$-string of radius $R$ is $2\pi R$.
Let $n_{\!_J}$ be the total number of line-elements in a $J$-string.
Thus
$
{2\pi R}
= 
n_{\!_J} {\ell_{\Delta}}
$.
The minimal possible value of the length $2\pi R$ must correspond 
to $n_{\!_J} = 1$. Let us denote the~radius of such a circumference 
by~$r_{\Delta}$. Explicitly, when $n_{\!_J} = 1$, we have 
$R = r_{\Delta}$ and $2\pi R = \ell_{\Delta}$.\pagebreak[3.99]
For the foregoing sequence of $J$-strings of 
decreasing radii, assuming that the number of elements
in this sequence is unbounded, we have $R_i \to r_{\Delta}$
in the limit.

According to~\eq{e5},
the radius of a~$J$-string and its mass
are related to each other in inverse proportion. Thus if there 
exists a limiting value for the radius of a~$J$-string, the
maximum possible mass of a~$J$-string is also limited.
Let us denote this mass by $m_{\Delta}$, and observe that,
as the mass of a~$J$-string increases, the limit of the ratio
of its charge to its mass is $q_{\!_J}/m_{\Delta}$.
Therefore, we conclude that the limiting value
$\varphi_z^*$ introduced in~\S\ref{6.2.2} 
can be described by the following expression 
\begin{equation}
	\label{e28a}
	\varphi_z^* 
	\ = \
	\left(\frac{q_{\!_J}}{m_{\Delta}}\right)^2
\end{equation}

\subsubsection{}\label{4.1.3}
Let us write Newton's law in two forms
\begin{equation*}
	F_g = G\frac{m\cdot m }{L^2}; \quad 
	F_g = \frac{(\sqrt{G}m) (\sqrt{G}m)}{L^2}.
\end{equation*}

\subsubsection{}
As is known, the expression $\sqrt{G}\cdot m$ 
has the dimension of charge. 
Let us denote this charge by $q_g$; we obtain
\begin{equation}
	F_g = G\frac{m\cdot m }{L^2} = \frac{q_g\cdot q_g}{L^2} = \frac{q^2_g }{L^2}
\end{equation}
which means
\begin{equation}\label{e29aa}
	\left(\frac{q_g }{m}\right)^2 = G.
\end{equation}
Let us now turn to $J$-strings;
we assume that $q_{\!_J} = \sqrt{G}m_\Delta$, i.e., formula~\eq{e29aa}
is also valid for $\varphi_z^*$.
It follows that
\begin{equation}\label{e29b}
	\varphi_z^*=\left(\frac{q_{\!_J}}{m_{\Delta}}\right)^2=G.
\end{equation}

\subsubsection{}
Based upon~\eq{e29b}, the maximal mass $m_{\Delta}$ can 
be determined using~\eq{e17} as follows 
\begin{equation}
	\label{e30}
	m_{\Delta}
	\ = \ 
	\frac{q_{\!_J}}{\sqrt{G}}
	\ = \ 
	\left({\frac{\hbar c}{G}}\right)^{1\over2}
\end{equation}
(The general form of the foregoing expression for $m_{\Delta}$ 
indicates that the maximal mass $m_{\Delta}$ constitutes one\pagebreak[3.99] 
of the ``Planck units;'' it is assumed that $m_{\Delta}$
can be regarded as the Planck mass $m_P$.)
From \eq{e30} and~\eq{e4}, we can 
now determine the value of $r_{\Delta}$,
that is, the fundamental length $A$. 
Namely: 
\begin{equation}\label{e31}
	A
	\ = \ 
	r_{\Delta}
	\ = \
	\left({\frac{\hbar G}{c^3}}\right)^{1\over2}
	\vspace{1.0ex}
\end{equation}	
The fundamental length $B$ is,
correspondingly, given by\vspace{1ex}
\begin{equation}\label{e32}
	B
	\ = \ 
	\ell_{\Delta}
	\ = \
	2\pi \left({\frac{\hbar G}{c^3}}\right)^{1\over2}
	\vspace{1.25ex}
\end{equation}

\subsubsection{}	 
The values of the physical quantities described above are:
\begin{eqnarray*}
	m_\Delta    & = & m_P \ \ \hspace{1.1ex} = 
	\ 2.1767 \cdot 10^{-8}\, {\rm kg}\\
	r_\Delta    & = & (\ell_P)_1 \ = \ 1.616 \cdot 10^{-35}\, {\rm m}\\
	\ell_\Delta & = & (\ell_P)_2 \ = \ 1.015 \cdot 10^{-34}\, {\rm m}
\end{eqnarray*}


\sadvance
\subsubsection{}\label{6.3.1}
A physical line-element must have constant curvature
along its entire extent of length~$\ell_{\Delta}$.
To establish this, assume to the contrary that
the curvature of a certain physical line-element 
varies smoothly, from curvature $1/R_a$ 
at one end 
to curvature $1/R_b$ at the other end. 

\looseness=-1
Consider a continuous line, whose curvature varies smoothly
over a given section (which can be as small as desired) of length $\ell$.
As is well known, such a line can be mathematically represented as 
a sequence of $n$ segments, each of length $d\ell$, where $n \to \infty$
and $d\ell \to 0$. All the segments of length $d\ell$ have different 
curvature. However, along each such segment 
the curvature is constant: $1/R = \text{const}$.
If this approach is applied to the line-segment
of length $\ell_{\Delta}$, we obtain a sequence 
of $n^*$ segments: 
$(d\ell_{\Delta})_1$ of constant curvature $1/R_1$, 
$(d\ell_{\Delta})_2$ of constant curvature $1/R_2$, 
and so forth.

\looseness=-1
Consequently, 
one can distinguish within
the small line segment of length $\ell_{\Delta}$ even smaller segments
of length $\ell^*_{\Delta}$ that differ from each other in their 
curvature. But this is impossible. Since $\ell_{\Delta}$ is the 
minimum ``unit of length,'' indivisible into parts,
distinguishing, or telling apart, any curvilinear segments of 
length $\ell^*_{\Delta} < \ell_{\Delta}$ is impossible in principle.
Hence, along the entire extent of length $\ell_{\Delta}$ of the 
line-element --- 
as well as along the entire extent of a~segment $d\ell$ of 
any known continuous curve --- we have $1/R = \text{const}$.
(Instead of the foregoing assumption that the curvature of
a physical line-element {\sl varies smoothly\/} from $1/R_a$
to $1/R_b$, one could also consider the assumption of 
{\sl discrete variation\/} of its curvature; since any
radius is equal to a multiple $nr_{\Delta}$ of $r_{\Delta}$,
we have $R_a = n_a r_{\Delta}$ and $R_b = n_b r_{\Delta}$.
Using this approach, we come to the same conclusion that 
along the entire extent of the length $\ell_{\Delta}$, the 
value of $1/R$ must be constant.)

\subsubsection{}\label{6.3.2}
\looseness=-1
Based on the foregoing analysis,
we arrive at the following conclusion:
quantum lines (i.e., lines consisting
of line-elements) are fundamentally different from 
conventional (non-quantum) continuous lines;
one such fundamental difference is the
{\sl impossibility of having an arbitrarily chosen form}.

Any part of a physical (quantum) line is composed of line-elements 
of a planar, and only planar, 
line. A continuous physical (quantum) line cannot be  
a spatial curve, such as a helix. 
(The existence of any quantum spatial curve hinges upon 
the existence of a line-element having both curvature and torsion;
hence one has to assume the existence of yet another, third,
fundamental length --- in addition to the two ``units of length''
of planar lines, introduced in~\S\ref{4.1.2real}\,--\,\S\ref{6.1.4}.)

The condition requiring constant curvature along each line-element 
implies that 
{\sl the only possible form of a planar curve of length $L$\/}
is a circle (or part of a circle).
Any other form is ruled out by this condition. 
Thus, for example, a quantum curve ``in the form of an ellipse''
would be composed of line-elements of circles of different radii, 
each having length $\ell_\Delta$; these line-elements would be 
sequentially conjoined to each other to form the ``line of the 
ellipse'' but with cusp singularities at all such points of juncture.
\pagebreak[3.99]

\vspace{12.5ex}
\section{Conclusions}
\label{sec5} 
\sreset

\subsubsection{} 
This article presents a body 
of interrelated physical concepts 
and ideas. We have introduced and studied certain 
linear objects --- {physical lines}, which have been
interpreted as strings. A~closed planar line of constant curvature
was investigated in-depth.
The term {$J$-string\/}
was introduced to denote this physical curve.

\subsubsection{} 
\looseness=-1
It was assumed that a~$J$-string has an angular momentum of $\hbar$.
It was then demonstrated that the $J$-string rotates
in its plane about its geometric center with linear~velocity~$c$.

\subsubsection{} 
It was established that a~$J$-string can exist in three distinct states
$\alpha$, $\beta$, and $\gamma$; under certain conditions, it may transit
from one state to another. In state $\alpha$, a~$J$-string is expanding 
(contracting) concentrically with velocity $c$, while its geometric 
center is at rest. In state~$\beta$, \mbox{a~$J$-string} is moving 
at velocity $c$ along an axis that is perpendicular to its plane,
while in state~$\gamma$ it can move (displace) at any velocity $v$,
provided $0 \le v < c$.

\subsubsection{} 
\looseness=-1
Physical properties and characteristics of a $J$-string of radius $R$
in state $\gamma$ have been~investigated.
It was shown that such a $J$-string (as well as all $J$-strings) 
possesses a~mass, an~energy, and a~charge $q_{\!_J}$. The mass
$m_{\!_J}$ of a~$J$-string can be expressed as $b/R$, 
where $b = h/2\pi c$.
The charge $q_{\!_J}$ is equal to $({hc/2\pi})^{1/2}$.
The circumference line of a~$J$-string is
uniformly distended along its entire length by a force directed along 
the radii, whose absolute value~is~$hc/2\pi R^2$.

\sadvance
\subsubsection{} 
It was assumed that the circumference line of a~$J$-string is composed 
of elements of a~certain length.
It was established that the minimal possible radius of a~$J$-string
can be expressed as~$({\hbar G/c^3})^{1/2}$, while the length 
of its line-element is given by $2\pi({\hbar G/c^3})^{1/2}$.

\subsubsection{} 
Various aspects of the problem concerning the Planck length 
have been studied. As is well known, the Planck length is
regarded in string theory as a fundamental length, which
determines the dimensions of a quantum of space.

\looseness=-1
Based upon investigation of the properties and characteristics
of $J$-strings, a qualitatively new method for the computation
of the fundamental length was developed. 
It was shown that one has to consider, not one, but 
two primary linear parameters: $(\ell_P)_1 = r_\Delta$ 
and $(\ell_P)_2 = \ell_\Delta$.

\subsubsection{} 
\looseness=-1
The problem of the ``maximal mass'' has been investigated.
It is well known that the expression $(hc/2\pi G)^{1/2}$ 
serves as the basis for the computation of the Planck 
mass $m_P$ --- a~certain ``maximal mass.'' 
It is widely assumed in the literature
that this mass can be correlated with 
the minimal volume~$\ell_P^3$. Thus the expression 
$m_P/\ell_P^3$ is usually invoked in estimating 
the value of the ultimate density (this parameter 
plays a fundamental role in the theories of gravity 
currently being developed). 

\looseness=-1
The question of the existence of a ``maximal mass''
is extremely important. 
We have
obtained an expression for the mass of a hypothetical
object (a $J$-string), as well as a~formula for the 
computation of the maximal (maximum possible) mass
of an elementary object --- the mass $m_{\Delta}$.
Specifically $m_{\Delta} = (\hbar c/G)^{1/2}$
(it is assumed that $m_\Delta$ is, in fact, 
the mass~$m_P$).

\subsubsection{} 
It was established that with the maximum (limiting possible) 
mass of an elementary object there is uniquely associated 
a certain physical quantity that may be interpreted as charge. 
This quantity was denoted by $q_{\!_J}$. 
The charge $q_{\!_J}$ is equal to $(m_P G)^{\frac{1}{2}}$, 
and $q_{\!_J}^2 \cong 137 e^2$. 
Since the charge $q_{\!_J}$ is described by the expression $\left(\frac{ h c}{2\pi}  \right)^{\frac{1}{2}}$, 
it appears appropriate to regard it --- together 
with the fundamental length and the maximum mass --- as one of the ``Planck quantities''.

\subsubsection{} 
The notion of strings has been used in elementary 
particle physics (in hadron models) since the late 
1960's. In the middle 1970's, upon publication of
the well-known work of Scherk and Schwarz~\cite{ScSc},
this notion has been developed in a qualitatively new 
way: strings have come to be regarded as objects that
could serve as the physical basis for a~unified
description of electromagnetic, weak, and strong 
interactions, and gravitation. In subsequent decades,
numerous authors have introduced and studied different
models of strings, in an attempt to construct such a~unified
theory. As is well known, the efforts to solve
this problem have invariably run into more and more
difficulties; this indicates that one might need to
look for fundamentally new directions in the investigation
of strings.

\subsubsection{}
In the development of all the existing models of strings
(as well as in the earlier hadron models), the following
underlying assumption was adopted: a ``basic string'' must possess
certain specific properties corresponding to the 
properties of the actual fields (particles) that are
described with the help of this string model.
However, a completely different approach to the construction
of the ``basic string'' model is also conceivable: refrain from 
considering the properties of the existing fields (particles)
and consider instead the properties of vacuum --- the primary,
most fundamental physical entity. The present work is an attempt
to develop one such model --- the {\sl $J$-string}. This model
is constructed on the basis of a certain system of axioms
pertaining to vacuum; it is thus the properties of vacuum 
that are reflected in the properties of a $J$-string.

\vspace{12ex}

{\small

}

\end{document}